\documentclass[12pt]{article}
\usepackage[dvips]{epsfig}

\begin{document}

\author{A. de Souza Dutra$^{a,b}$\thanks{%
e-mail: dutra@feg.unesp.br}, M. Hott$^{b}$\thanks{%
e-mail: hott@feg.unesp.br} and F. A. Barone$^{b}$\thanks{%
e-mail: fbarone@feg.unesp.br} \\
$^{a}$Abdus Salam ICTP, Strada Costiera 11, Trieste, I-34100 Italy.\\
$^{b}$UNESP-Campus de Guaratinguet\'{a}-DFQ\thanks{ Permanent
Institution}\\
Departamento de F\'{\i}sica e Qu\'{\i}mica\\
12516-410 Guaratinguet\'{a} SP Brasil}
\title{{\LARGE Two Field BPS Solutions for Generalized Lorentz Breaking Models}}
\maketitle

\begin{abstract}
In this work we present non-linear models in two-dimensional
space-time of two interacting scalar fields in the Lorentz and CPT
violating scenarios. We discuss the soliton solutions for these
models as well as the question of stability for them. This is done
by generalizing a model recently published by Barreto and
collaborators and also by getting new solutions for the model
introduced by them.

\end{abstract}

\newpage

\smallskip Some years ago, Carrol et al \cite{Carrol} start to analyze the
problem of Lorentz breaking signature in field theoretical models. By now,
there are a great number of works discussing a kind of symmetry breaking in
many different physically interesting contexts. For instance, in \cite%
{Colladay} it was discussed some impact over the standard model of
this kind of symmetry breaking. Azatov {\it et al} \cite{Azatov}, in a
recent work, have analyzed the spontaneous breaking of the four
dimensional Lorentz invariance of the QED through a nonlinear
vector potential constraint, Bezerra {\it et al} \cite{Bezerra} have
shown that a space-time with torsion interacting with a Maxwell
field by means of a Chern-Simons like term can explain the optical
activity in the sinchroton radiation emitted by cosmological
distant radio sources, Lehnert {\it et al} \cite{Lehnert} have verified
the consequences over the Cerenkov effect of a Lorentz-violating
vacuum and Bluhm \cite{Bluhm} has made an estimative analysis of
the Lorentz and CPT bounds attainable in Penning-trap experiments.
In fact, along the last years a considerable effort has been drawn
into this direction by many groups and in a variety of physical
applications. On the other hand, the presence of topological
solutions of nonlinear models is a matter of large interest and
possible applications \cite{Rajaraman,Vilenkin,Walgraef}. On the
other hand, a natural place to apply these ideas is that of
condensed matter non-relativistic ground, where the break of
isotropy and homogeneity emerges quite naturally, due the material
structure.

As a consequence of the above arguments, it is natural to look for
topological structures in CPT breaking scenarios. In fact, in a very recent
work in this journal, Barreto {\it et al} \cite{Barreto} have introduced an
approach capable of getting kinks in CPT violating scenarios.

Here we are going to discuss a generalization of the work of reference \cite%
{Barreto}, both by obtaining more general solutions for the models
considered on that work and by generalizing Lorentz breaking
Lagrangian densities.
Particularly we obtain solutions which were absent in the reference \cite%
{Barreto}. For this last accomplishment, we use a method recently introduced
by one of us \cite{PLB05}.

Models with Lorentz breaking terms usually leads to non-linear differential
equations, and one of the problems appearing as a consequence of this
nonlinearity is that, in general, we loose the capability of getting the
complete solutions. Here we extend an approach exposed in reference \cite%
{PLB05} which shows that for some two field systems in 1+1 dimensions, whose
the second-order differential equations can be reduced to the solution of
corresponding first-order equations (the so called
Bolgomol'nyi-Prasad-Sommerfield (BPS) topological solitons \cite{BPS}), one
can obtain a differential equation relating the two coupled fields which,
once solved, leads to the general orbit connecting the vacua of the model.
In fact, the \textquotedblleft trial and error\textquotedblright\ methods
historically arose as a consequence of the intrinsic difficulty of getting
general methods of solution for nonlinear differential equations. About two
decades ago, Rajaraman \cite{Rajaraman} introduced an approach of this
nature for the treatment of coupled relativistic scalar field theories in
1+1 dimensions. His procedure was model independent and could be used for
the search of solutions in arbitrary coupled scalar models in 1+1
dimensions. However, the method is limited in terms of the generality of the
solutions obtained and is convenient and profitable only for some
particular, but important, cases \cite{boya}. Some years later, Bazeia and
collaborators \cite{bazeia0} applied the approach developed by Rajaraman to
special cases where the solution of the nonlinear second-order differential
equations are equivalent to the solution of corresponding first-order
nonlinear coupled differential equations. In this work we are going to
present a procedure which is absolutely general when applied to Lorentz and
CPT breaking systems, like those obtained from an extension of the ones
described in \cite{PLB05} applied to nonbreaking versions appeared in \cite%
{bazeia0}-\cite{bazeia4}. Furthermore, we also show that many of these
systems can be mapped into a first-order linear differential equation and,
as a consequence, can be solved in order to get the general solution of the
system. After that, we trace some comments about the consequences coming
from these general solutions.

\section{BPS nonlinear Lorentz and CPT scenarios}

The two field model we shall study in $1+1$ dimensions is described by the
Lagrangian density
\begin{equation}
\mathcal{L}=\frac{1}{2}(\partial _{\mu }\phi
)^{2}+\frac{1}{2}(\partial _{\mu }\chi )^{2}-f^{\mu }(\phi ,\chi
)\partial _{\mu }\chi -g^{\nu }(\phi ,\chi )\partial _{\nu }\phi
-V(\phi ,\chi ),  \label{calL}
\end{equation}
where $\mu =0,1$, $f^{\mu }(\phi ,\chi )$ and $g^{\nu }(\phi ,\chi )$ are
vector functions with a prescribed functional dependence on the dynamical
fields $\phi $ and $\chi $, and $V(\phi ,\chi )$ is a potential term.

Note that we can recover some usual Lorentz symmetry breaking models from (%
\ref{calL}) by choosing appropriately the vectors $f^{\mu }$ and $g^{\mu }$.
In particular, if $f_{1}=s_{2}\,\phi $ and $g_{1}=s_{1}\,\chi $, one recovers
the model introduced very recently by Barreto and collaborators \cite%
{Barreto}. In fact, the first example we work out here is
precisely this one, which we are going to show possesses an entire
topological sector not considered in the work of reference
\cite{Barreto}.

If the potential $V(\phi ,\chi )$ can be written in such a way that
\begin{equation}
V(\phi ,\chi )=\frac{1}{2}\Biggl(\frac{dW(\phi ,\chi )}{d\phi }-g_{1}(\phi
,\chi )\Biggr)^{2}+\frac{1}{2}\Biggl(\frac{dW(\phi ,\chi )}{d\chi }%
-f_{1}(\phi ,\chi )\Biggr)^{2}\ ,
\end{equation}
with $W(\phi ,\chi )$ being any function of $\phi $ and $\chi $, the energy
density of the BPS states becomes
\begin{equation}
\mathcal{E}_{BPS}=\frac{1}{2}\Biggl(\frac{d\phi }{dx}-\frac{dW(\phi ,\chi )}{
d\phi }+g_{1}(\phi ,\chi )\Biggr)^{2}+\frac{1}{2}\Biggl(\frac{d\chi }{dx}-
\frac{dW(\phi ,\chi )}{d\chi }+f_{1}(\phi ,\chi )\Biggr)^{2}\ +\frac{dW}{dx},
\label{energia}
\end{equation}
with $dW/dx=W_{\phi }\phi ^{\prime }+W_{\chi }\chi ^{\prime }$, where we have defined $W_{\phi }\equiv \frac{\partial W}{\partial \phi }$, $W_{\chi }\equiv \frac{\partial W}{\partial \chi }$ and the prime stands for
space derivative.

From equation (\ref{energia}), we can see that the solutions of minimal energy are obtained from the
following two coupled first order equations
\begin{eqnarray}
\phi ^{\prime } &=&W_{\phi }(\phi ,\chi )-g_{1}(\phi ,\chi )\ ,  \nonumber \\
&&  \label{eq1a} \\
\chi ^{\prime } &=&W_{\chi }(\phi ,\chi )-f_{1}(\phi ,\chi )\ ,  \nonumber
\end{eqnarray}

Finally the BPS energy is written, as usual, by

\begin{equation}
E_{BPS}=|W\left( \phi _{j},\chi _{j}\right) -W\left( \phi _{i},\chi
_{i}\right) |\ ,  \label{Ebps}
\end{equation}

\noindent where $\phi _{i}$ and $\chi _{i}$ mean the $i-th$ vacuum states of
the model. Here, it is important to remark that the BPS solutions settle
into vacuum states asymptotically. In other words, the vacuum states act as
implicit boundary conditions of the BPS equations.

It is interesting to notice that in the first order equations of motion (\ref%
{eq1a}) and in the energy density (\ref{energia}) only the space components
of the functional vectors $f_{\mu }$ and $g_{\mu }$, $f_{1}$ and $g_{1}$
respectively, are present.

From now on, in order to solve the equations (\ref{eq1a}), let us consider
models for which we can write $\phi$ as a function of $\chi$, that is, $\phi
(\chi)$. In this situation, instead of applying the usual trial-orbit
approach \cite{bazeia0}-\cite{bazeia4}, we note that it is possible to write
the following equation
\begin{equation}
\frac{d\phi }{W_{\phi }-g_{1}}=dx=\frac{d\chi }{W_{\chi }-f_{1}},
\end{equation}

\noindent where the differential element $dx$ is a kind of invariant. In
these cases one is lead to

\begin{equation}
\frac{d\phi }{d\chi }=\frac{W_{\phi }-g_{1}}{W_{\chi }-f_{1}}.  \label{eqm}
\end{equation}

Equation (\ref{eqm}) is the generalization of the one studied in \cite{PLB05}
to the case of nonlinear Lorentz and CPT breaking scenarios. It is, in
general, a nonlinear differential equation relating the scalar fields of the
model. If one is able to solve it completely for a given model, the function
$\phi \left( \chi \right)$ can be used to eliminate one of the fields, so
rendering the equations (\ref{eq1a}) uncoupled and equivalent to a single
one. Finally, this uncoupled first-order nonlinear equation can be solved in
general, even if numerically.

We have found this method simpler than the method of the orbits
broadly and successfully applied to study the mapping of the
soliton solutions and defect structures in problems involving the
interaction two scalar fields. Despite of being simpler, the
method applied here furnishes not only the same orbits than those
obtained by using the method of the orbits appearing in the
references \cite{bazeia0}-\cite{bazeia4}, but also some new ones
as can be seen through a comparison with reference \cite{PLB05}.
In the example worked out below one can verify that, this time,
the mapping constructed here furnishes the very same orbits
obtained in the reference \cite{Barreto}. Notwithstanding, we were
able to find new solitonic configurations, not observed by
Barreto and collaborators.

\section{The example of linear Lorentz and CPT breaking}

In this section we consider the particular model introduced in the work of
Barreto {\it et al} \cite{Barreto} in order to apply the method discussed in the
previous section. In fact, we show in this example that the equation (\ref%
{eqm}) can be mapped into a linear differential equation, from which it is
possible to obtain the general solutions for the soliton fields. In the case
on the screen, the superpotential \cite{Barreto} is written as
\begin{equation}
W\left( \phi ,\chi \right) =\,\phi -\frac{1}{3}\phi ^{3}-r\,\phi \,\chi ^{2},
\label{w1}
\end{equation}

\noindent and the Lorentz symmetry breaking terms in the lagrangian density (%
\ref{calL}) are chosen to be given by $f_{1}(\phi,\xi)=s_{2}\,\phi$ and $%
g_{1}(\phi,\xi)=s_{1}\,\chi$, such that equation (\ref{eqm}) is
rewritten as
\begin{equation}
\frac{d\phi }{d\chi }=\frac{\left( \phi ^{2}-1\right) +r\,\,\chi
^{2}+s_{1}\,\chi }{\,2\,r\,\phi \,\chi +s_{2}\,\,\phi }\ ,
\label{phi/chi}
\end{equation}
where $s_{1}$ and $s_{2}$ are constants.

At this point one can verify that, performing the transformations
\begin{equation}
\chi =\zeta -\frac{s_{2}}{2\,r},  \label{10}
\end{equation}%
and
\begin{equation}
\phi ^{2}=\rho +1+\frac{s_{2}}{4\,r}\left( 2\,s_{1}-s_{2}\right) ,
\label{10a}
\end{equation}%
the equation (\ref{phi/chi}) becomes
\begin{equation}
\frac{d\rho }{d\zeta }-\frac{\rho }{r\,\zeta }\,=\,\zeta \,-\frac{b}{r},
\label{11}
\end{equation}%

\noindent which is a typical inhomogeneous linear differential
equation \cite{PLB05}. The general solutions for the orbit
equation are then easily obtained, giving
\begin{equation}
\phi ^{2}-1=c_{0}\,\zeta ^{\frac{1}{r}}+\frac{r}{2\,r-1}\,\,\zeta ^{2}-\frac{%
b\,}{r-1}\,\zeta +k\ \qquad \mathrm{for}\ r\neq 1\ {\textrm{and}}\
r\neq \frac{1}{2},  \label{12a}
\end{equation}%
\begin{equation}
\phi ^{2}-1=-b~\zeta \,\ln \left( \zeta \right) +c_{1}\,\zeta +\zeta
^{2}+k,\qquad \mathrm{for}\ r=1  \label{12b}
\end{equation}%
\noindent and
\begin{equation}
\phi ^{2}-1=\zeta ^{2}\ln \left( \zeta \right) +b~\zeta +c_{2}\,\zeta
^{2}+k,\qquad \mathrm{for}\ r=\frac{1}{2},  \label{12c}
\end{equation}

\noindent where $k\equiv \frac{s_{2}}{4\,r}\left( 2\,s_{1}-s_{2}\right) $, $%
b\equiv s_{2}-s_{1}$ and $c_{0}$, $c_{1}$ and $c_{2}$ are arbitrary
integration constants.

In general it is not possible to solve $\chi $ in terms of $\phi$ from the
above solutions, but the contrary is always granted. Here, with the aid of (%
\ref{w1}) and (\ref{10}), we shall substitute the expressions of $\phi
\left(\chi \right)$ obtained from (\ref{12a}), (\ref{12b}) and (\ref{12c})
in the second equation (\ref{eq1a}), obtaining respectively:
\begin{equation}
\frac{d\zeta }{dx}=\pm \,2\,r\,\zeta \sqrt{\,1+c_{0}\,\zeta ^{\frac{1}{r}}+
\frac{r}{2\,r-1}\,\,\zeta ^{2}-\frac{\,b}{r-1}\,\zeta +k}\,,\quad \,\,{%
\textrm{for}}\ r\neq 1,~r\neq \frac{1}{2},  \label{13}
\end{equation}

\noindent\
\begin{eqnarray}
\frac{d\zeta }{dx}&=&\pm \,2\,r\,\,\zeta \,\sqrt{1-b\,\zeta \,\ln \left(
\zeta \right)+c_{1}\,\zeta +\zeta^{2}+k}\,,\quad \mathrm{for}\ r=1,
\nonumber \\
&& \\
\frac{d\zeta }{dx} &=&\pm \,2\ r\,\zeta \,\sqrt{1+\zeta ^{2}\,\ln
\left(\zeta \right) +b\ \zeta +c_{2}\ \zeta ^{2}+k}\ ,\quad \mathrm{for}\ r=%
\frac{1}{2}.  \nonumber
\end{eqnarray}

Barreto and collaborators \cite{Barreto} have limited themselves to the
orbits in which $r\neq 1$ and $r\neq 1/2$ and the arbitrary constant $c_{0}$
equals to zero or infinity. In the particular
case with $c_{0}=0$ they have found a lump-like profile for the field $\chi
(x)$ and a kink-like profile for the field $\phi (x)$. By integrating the
equation (\ref{13}) and substituting its solutions into the equation (\ref%
{10}) we get the following solutions for the field $\chi (x)$

\begin{equation}
\chi _{\pm }^{A}(x)=\frac{4\sqrt[3]{A}e^{\mp 2\sqrt{A}r(x-x_{0})}}{(\sqrt{A}%
e^{\mp 2\sqrt{A}r(x-x_{0})}+C)^{2}-4AB}-\frac{b}{2r},  \label{chiA}
\end{equation}
where $x_{0}$ is a constant of integration, $A=1-b^{2}/4r$, $B=r/(2r-1)$, $%
C=b/(r-1)$ and we have taken $s_{1}=0$. On its turn the solutions for the
field $\phi (x)$ are obtained by substituting the classical solutions of the
equation (\ref{13}) into the equation (\ref{12a}), namely

\begin{equation}
\phi _{\pm }^{A}(x)=\pm \frac{\sqrt{A}[Ae^{\mp 4\sqrt{A}%
r(x-x_{0})}-(C^{2}-4AB)]}{(\sqrt{A}e^{\mp 2\sqrt{A}r(x-x_{0})}+C)^{2}-4AB}\ .
\label{phiA}
\end{equation}

The above solutions are valid if the parameters satisfy the
conditions $A>0$ and $C^{2}\neq 4AB$. The behavior of the above
solutions are plotted in the figure 1 for the parameters $r=0.4$
and $b=0.6$. One can observe that in both pairs of solutions,
$(\phi _{+},\chi _{+})$ and $(\phi _{-},\chi _{-})$, the field
$\chi (x)$ exhibits a lump-like profile and the field $\phi (x)$ a
kink-like profile. This behavior is also found in many systems of
two interacting solitons reported in the literature.

More recently \cite{PLB05} it has been shown that many models of
two interacting solitons, very similar to this one with explicit
Lorentz symmetry breaking that we are presenting here, can also
exhibit kink-like behavior for both of the soliton fields,
depending on the range of the parameters of the model. Inspired on
this achievement, we have shown that it is also possible to have
kink-like profiles for both of the fields, for particular values
of the
parameters $r$ and $b$, in the model treated here. In fact if one takes $%
b=2(r-1)/\sqrt{r}$, which corresponds to one of the solutions with $%
C^{2}=4AB $, and $r>1/2$ in the equations (\ref{chiA}) and (\ref{phiA}) we
obtain the following forms for the fields

\begin{equation}
\chi _{\pm }^{B}(x)=\frac{4(2r-1)}{r\left( \sqrt{2r-1}~e^{\mp 2\sqrt{2r-1}%
(x-x_{0})}+4\sqrt{r}\right) }-\frac{r-1}{r\sqrt{r}},  \label{chiB}
\end{equation}
and

\begin{equation}
\phi _{\pm }^{B}(x)=\pm \frac{4(2r-1)}{r\left( \sqrt{2r-1}~e^{\pm 2\sqrt{2r-1%
}(x-x_{0})}+\sqrt{2r-1}\right)}\ .  \label{phiB}
\end{equation}
In the figure 2 we present the behavior of the above kink solutions for $r=2$%
.

One could interpret these solutions as representing two kinds of torsion in
a chain, represented through an orthogonal set of coordinates $\phi $ and $%
\chi $. So that, in the plane ($\phi $,$\chi $), the type-$A$ kink
corresponds to a complete torsion and the type-$B$ corresponds
to a half torsion, similarly to what has been done in \cite{PLB05}.

It is worth mentioning that the pairs of type-$B$ solutions have a BPS energy
lower than that associated to the type-$A$ soliton solutions. This can be
shown by substituting the asymptotic values of the solutions in the equation
(\ref{Ebps}), that is, for the type-$A$ solutions we find $E_{BPS}^{A}=\frac{4%
}{3}A\sqrt{A}$, and $E_{BPS}^{B}=\frac{2}{3}A\sqrt{A}$ for the type-$B$
solutions.

\section{Generalized models}

In what follows, we will study a more general model contemplating a number
of particular cases which have been studied in the literature, including the
previous and some other new ones. For this, we begin by defining the
superpotential
\begin{equation}
W\left( \phi ,\chi \right) =\frac{\mu }{2}\,\phi ^{N}\,\chi ^{2}+F\left(\phi
\right)\ ,
\end{equation}
such that the equation (\ref{eqm}) is given by

\begin{equation}
\frac{d\phi }{d\chi }=\frac{F_{\phi }+\frac{\mu }{2}\,N\,\,\phi ^{\left(
N-1\right) }\,\chi ^{2}-g_{1}(\phi ,\chi )}{\mu N\,\,\phi ^{N}\,\chi
-f_{1}(\phi ,\chi )},  \label{phi/chi-2}
\end{equation}
where $F_{\phi}=dF/d\phi$. The space-component of the functionals terms
responsible for breaking the Lorentz symmetry explicitly, namely, $%
f_{1}(\phi ,\chi )$ and $g_{1}(\phi ,\chi )$, are to be chosen more general
than those of the model discussed previously and conveniently such that the
integration of the equation (\ref{phi/chi-2}) be possible. Based on the
succesfull generalization of models of interacting solitons also carried out
in the reference \cite{PLB05} and in the development of the model of the
previous section, a possible generalized model can be constructed by choosing

\begin{equation}
F(\phi )=\frac{1}{2}\phi ^{N}\left( \frac{\lambda }{N+2}\phi ^{2}+\frac{%
\gamma }{N}\right)\ ,  \label{F(phi)}
\end{equation}
and the following forms for the functionals $f_{1}(\phi,\chi)$ and $%
g_{1}(\phi,\chi)$,

\begin{eqnarray}
f_{1}(\phi ,\chi ) &=&b\,\phi ^{N}\chi\ ,  \nonumber \\
g_{1}(\phi ,\chi ) &=&a\,\phi ^{N-1}\chi\ ,
\end{eqnarray}

\noindent where $N$ is a positive integer number, $\lambda$ and $\gamma$ are
constants and the parameters $a$ and $b$ can be thought as space-components
of two-vectors pointing out in some preferred direction in space-time and
the responsible for breaking the Lorentz symmetry.

The corresponding equation for the dependence of the field $\phi $ as a
function of the field $\chi$ is now given by
\begin{equation}
\frac{d\phi }{d\chi }=\frac{1}{2}\frac{\mu \,N\,\,\phi ^{N-1}\,\chi
^{2}+\phi ^{N-1}(\lambda \phi ^{2}+\gamma )-2a\phi ^{N-1}\chi }{\mu \,\phi
^{N}\,\chi -b\phi ^{N}}\ .
\end{equation}
Now, by performing the transformations

\begin{equation}
\sigma =\frac{1}{2\mu }\left( \lambda \phi ^{2}+\frac{Nb^{2}}{\mu }+\gamma -%
\frac{2ab}{\mu }\right)\ ,  \label{trans1}
\end{equation}
and

\begin{equation}
\varsigma =\mu \chi -b,  \label{trans2}
\end{equation}
we get

\begin{equation}
\frac{d\sigma }{d\varsigma }-\frac{\lambda \sigma }{\mu \varsigma }=\frac{%
N\lambda }{2\mu ^{3}}\varsigma +\frac{\lambda }{\mu ^{3}}(Nb-a)\ .
\label{15}
\end{equation}

The above equation is very similar to the equation (\ref{11}) and can be
easily integrated out. Its general solution in the case $\lambda \neq \mu$
and $\lambda \neq 2\mu$ is

\begin{equation}
\sigma \left( \varsigma \right) =\frac{\lambda (Nb-a)}{\mu ^{2}(\mu -\lambda
)}\varsigma +\frac{N\,\lambda \,}{2\,\mu ^{2}\left( 2\mu -\lambda \right)}
\varsigma ^{2}+c\,\,\varsigma ^{\frac{\lambda }{\mu }},  \label{map2}
\end{equation}
where $c$ is an arbitrary integration constant. The solutions for the equation (\ref{15}%
) in the cases $\lambda =\mu $ and $\lambda =2\mu $ can also be obtained,
but we will not deal with them here.

We substitute the equations (\ref{trans1}), (\ref{trans2}) and (\ref{map2})
in one of the equations (\ref{eq1a}) to obtain the following first-order
equation of motion for the field $\varsigma$

\begin{equation}
\frac{d\varsigma }{dx}=\pm \mu ^{1-N/2}\varsigma \left[ \frac{N}{2\mu
-\lambda }\varsigma ^{2}+\frac{2(Nb-a)}{\mu -b}\varsigma +\frac{c\mu}{%
\lambda }\varsigma ^{\lambda /\mu }-Nb^{2}-\gamma \mu +2ab\right]^{N/2}.
\label{asd1}
\end{equation}

This last equation can be solved analytically or numerically,
depending on the values of the parameters. For the particular case
with $N=2$, $2b=a$ and $c=0 $ we obtain very simple kink solutions
for both of the fields $\phi (x)$ and $\chi(x)$, as can be
verified from the behavior of the solution for the field
$\varsigma(x)$

\begin{equation}
\varsigma (x)=\pm \frac{\sqrt{B}e^{Bx}}{\sqrt{1+Ae^{2Bx}}}\ ,
\label{varsigma}
\end{equation}
where $A=2/(2\mu -\lambda )>0$ and $B=(\gamma \mu -2b^{2})/\lambda >0$, and
by substituting (\ref{varsigma}) in the equations (\ref{trans2}), (\ref{map2}%
) and (\ref{trans1}).

The construction of an even more general model which includes non-linear
dependence on the field $\chi(x)$ can be carried out by following the
generalization proposed in the reference \cite{PLB05}. This can be
accomplished by choosing the following form of the superpotential

\begin{equation}
W_{NM}\left( \phi ,\chi \right) =\frac{\mu }{M}\,\phi ^{N}\,\chi
^{M}+F\left( \phi \right) ,  \label{superpot}
\end{equation}
where $F(\phi)$ is given by the equation (\ref{F(phi)}) and $M$ is a
positive integer. In order to include the terms responsible for breaking the
Lorentz symmetry and to obtain a solution for the differential equation (\ref%
{eqm}) it is reasonable to choose the functionals $f_{1}(\phi,\chi)$ and $%
g_{1}(\phi,\chi)$ in the following forms

\begin{eqnarray}
f_{1}(\phi ,\chi ) &=&b\,\phi ^{N}\chi ^{M-1}  \nonumber \\
g_{1}(\phi ,\chi ) &=&a\,\phi ^{N-1}\chi ^{M}\ .
\end{eqnarray}
With this generalization the equation (\ref{eqm}) can be written in the form

\begin{equation}
\frac{d\varphi}{d\chi}-\frac{\varphi}{(\mu-b)\lambda}\chi^{1-M}=\frac{(\mu
N-2aM)}{M(\mu-b)\lambda}\chi\ ,  \label{asd2}
\end{equation}
where $\varphi =\lambda \phi ^{2}+\gamma$.

The equation (\ref{asd2}) is similar to the one which appears in reference
\cite{PLB05}. It admits the solution

\begin{eqnarray}
\varphi(\chi)&=&\exp\Biggl[-\frac{1}{\lambda(\mu-b)}\frac{1}{M-2}\chi^{(2-M)}%
\Biggr]\times\cr\cr
&\ &\Bigg[{\tilde c}_{1}+\frac{2^{M/(M-2)}}{M(M-2)}\frac{\mu N-2aM}{%
2\lambda(\mu-b)}\chi^2 \Biggl(\frac{\chi^{(2-M)}}{(M-2)}\Biggr)%
^{2/(M-2)}\times\cr\cr
&\ &\ \ \ \ \ \ \ \ \Gamma\Biggl(\frac{2}{(M-2)},\frac{1}{\lambda(\mu-b)}%
\frac{\chi^{(2-M)}}{(M-2)}\Biggr)\Bigg]
\end{eqnarray}
where ${\tilde c}_{1}$ is an arbitrary integration constant and $\Gamma(a,z)=\int_{z}^{\infty}t^{a-1}e^{-t}dt$ is the incomplete Gamma function.

\section{Conclusions}

We have been able to generalize a model presented recently in the
reference \cite{Barreto} which incorporates the phenomena of
solitons interactions and the Lorentz symmetry breaking. The
generalization has been carried out in two ways. We have found
non-trivial classical solutions which exhibit kink-like behavior
for both of the interacting fields and, consequently, with BPS
energy lower than that associated with the usual solutions
presented previously for the same model. Another interesting
aspect of the kink-like solutions rest on the study of the
stability of the solutions against small time-dependent linear
perturbation. At least for some models with only one scalar field,
it has been shown in the reference \cite{Nami} that models with
kink-like solutions possess the stability of these solutions, on
the other hand, models with lump-like classical solutions are
unstable. For two interacting scalar fields the problem is
cumbersome,\ even though the authors of the reference
\cite{Barreto} have been able to show, based on very elegant and
general arguments, that the\ solutions found there, even with
lump-like configurations for one of the fields, are stable. We
understand that the analysis of the stability carried out in reference \cite{Barreto} is
valid for reference systems in which $b_{0}=0$, where $b_{0}$ is
the time-component of the two-vector responsible for the Lorentz
symmetry breaking. For reference systems in which $b_{0}\neq 0$
the analysis has not been done.

We have also proposed generalizations of the model of the reference \cite%
{Barreto} by introducing non-linear terms that break the Lorentz
symmetry. This last generalization was possible thanks to the
successful generalization carried out in the reference
\cite{PLB05} which deals with a Lorentz symmetric two-dimensional
model of interacting scalar fields.

\bigskip \bigskip \bigskip

\textbf{Acknowledgements: }The authors ASD and MH thanks to CNPq and FAB to
FAPESP for the financial support. We also thanks to Professor D. Bazeia for
introducting us to this matter and to the reference \cite{Barreto}. This
work has been finished during a visit of ASD within the Associate Scheme of
the Abdus Salam ICTP.

\newpage

\begin{figure}[tbp]
\begin{center}
\begin{minipage}{20\linewidth}
\epsfig{file=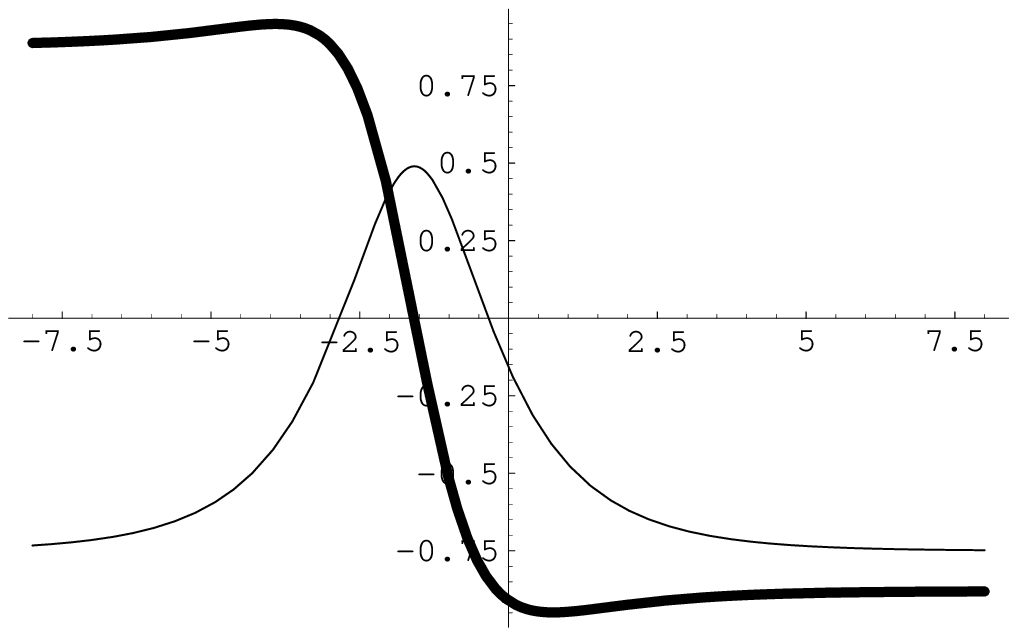}
\end{minipage}
\end{center}
\caption{Typical type-$A$ kink profile (for $r=0.6$, $b=0.4$). The
thin line corresponds to the field $\protect\chi_+(x)$ and the
thick line to the field $\protect\phi_+(x)$. Both were calculated
for $c_0=0$. } \label{fig:fig1}
\end{figure}

\begin{figure}[tbp]
\begin{center}
\begin{minipage}{20\linewidth}
\epsfig{file=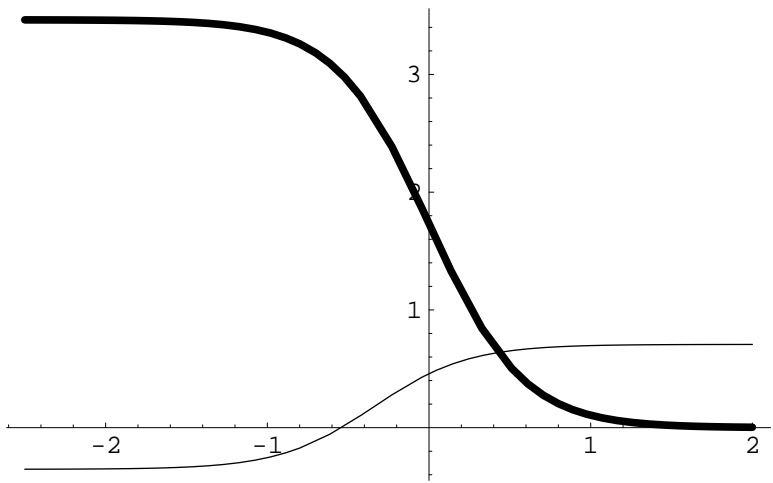}
\end{minipage}
\end{center}
\caption{Typical type-$B$ kink profile (for $r=2$). The thin line
corresponds to the field $\protect\chi_+(x)$ and the thick line to
the field $\protect\phi_+(x)$. Both were calculated for $c_0=0$.}
\label{fig:fig2}
\end{figure}


\newpage

\begin{thebibliography}{99}
\bibitem{Carrol} S. Carrol, G. Field and R. Jackiw, Phys. Rev. D \textbf{41}
(1990) 1231.

\bibitem{Colladay} D. Colladay and V. A. Kostelecky, Phys. Rev. D \textbf{55}
(1997) 6760; \textbf{58} (1998) 116002.

\bibitem{Azatov} A. I. Azatov and J. L. Chkareuli, Phys. Rev. D \textbf{73}
(2006) 065026.

\bibitem{Bezerra} V. B. Bezerra, C. N. Ferreira and J.A. Helayel-Neto, Phys.
Rev. D \textbf{71} (2005) 044018.

\bibitem{Lehnert} R. Lehnert and R. Potting, Phys. Rev. D \textbf{70} (2004)
125010.

\bibitem{Bluhm} R. Bluhm, Phys. Rev. D \textbf{57} (1998) 3932.

\bibitem{Rajaraman} R. Rajaraman, \textit{Solitons and Instantons}
(North-Holand, Amsterdam, 1982)

\bibitem{Vilenkin} A. Vilenkin and E. P. S. Shellard, \textit{Cosmic Strings
and Other Topological Defects} (Cambridge, Cambridge, UK, 1994).

\bibitem{Walgraef} D. Walgraef, \textit{Spatio-Temporal Pattern Formation}
(Springer-Verlag, New York, 1997).

\bibitem{Barreto} M. N. Barreto, D. Bazeia and R. Menezes, Phys. Rev. D
\textbf{73} (2006) 065015.

\bibitem{PLB05} A. de Souza Dutra, Phys. Lett. B \textbf{626} (2005) 249.

\bibitem{BPS} M. K. Prasad and C. M. Sommerfield, Phys. Rev. Lett. \textbf{35%
} (1975) 760; E. B. Bolgomol'nyi, Sov. J. Nucl. Phys. \textbf{24}
(1976) 449.

\bibitem{boya} L. J. Boya and J. Casahorran, Phys. Rev. A \textbf{39} (1989)
4298.

\bibitem{bazeia0} D. Bazeia, M. J. dos Santos and R. F. Ribeiro, Phys. Lett.
A \textbf{208} (1995) 84.

\bibitem{bazeia1} D. Bazeia and F. A. Brito, Phys. Rev. Lett. \textbf{84}
(2000) 1094.

\bibitem{bazeia1.5} D. Bazeia and F. A. Brito, Phys. Rev. D \textbf{61 }%
(2000) 105019.

\bibitem{bazeia2} D. Bazeia, R. F. Ribeiro and M. M. Santos, Phys. Rev. E
\textbf{54} (1996) 2943.

\bibitem{bazeia3} D. Bazeia and E. Ventura, Chem. Phys. Lett. \textbf{303}
(1999) 341.

\bibitem{bazeia4} D. Bazeia, V. B. P. Leite, B. H. B. Lima and F. Moraes,
Chem. Phys. Lett. \textbf{340} (2001) 205.

\bibitem{Nami} G. Flores-Hidalgo and N. F. Svaiter, Phys. Rev D \textbf{66}
(2002) 025031.
\end{thebibliography}
\end{document}